# Hubbard Models with Superconducting Quantum Symmetry

## Peter Schupp

Sektion Physik, Lehrstuhl Prof. Wess, Universität München
Theresienstr. 37, 80333 München, Germany

peter.schupp@physik.uni-muenchen.de

The current interest in the Hubbard model [1] is driven by the hope that it may aid in the understanding of high-$T_c$ superconductivity. Originally the model was introduced as a very simple description of narrow $d$-bands in transition metals.

The standard Hubbard Hamiltonian is

$$H_{\text{Hub}} = u \sum_i n_{i\uparrow} n_{i\downarrow} - \mu \sum_{i,\sigma} n_{i\sigma} + t \sum_{\langle i,j \rangle \sigma} a^\dagger_{j\sigma} a_{i\sigma}, \qquad (1)$$

where $a^\dagger_{i\sigma}$, $a_{i\sigma}$ are fermionic operators for electrons of spin $\sigma \in \{\uparrow, \downarrow\}$ at site $i$ of the lattice, $n_{i\sigma} \equiv a^\dagger_{i\sigma} a_{i\sigma}$, and $\langle i,j \rangle$ denotes nearest neighbor sites $i \neq j$. The model is given in grand canonical form; $\mu$ fixes the average number $\langle \sum_{i,\sigma} n_{i\sigma} \rangle$ of electrons. The first term is a local on-site interaction; for $u > 0$ it describes a repulsive Coulomb interaction, for $u < 0$ it yields an effective attractive interaction (*e.g.* mediated by ions). In the limit $u = 0$ the non-local *hopping-term* dominates and the model describes non-interacting moving electrons, *i.e.* band-like behavior, while in the opposite limit $t = 0$ the electrons are fully localized (atomic limit). The $u$- and $t$-terms are in competition with one another—this makes the model both very interesting and difficult. It is thus important to study symmetries of this and related models to obtain information about the spectrum and the possibility of off-diagonal long range order—see *e.g.* [2,3].

Since the work of Heilmann, Lieb, Yang and Zhang [4,2,5] the standard Hubbard model is known to have a SO(4) = SU(2) × SU(2)/$\mathbf{Z}_2$ symmetry at $\mu = u/2$, the value of $\mu$ corresponding to half filling in the band-like limit. This symmetry is the product of a *magnetic* SU(2)$_m$ (spin), which accounts for the (anti ferro-)magnetic properties of the electron system, and a *superconducting* SU(2)$_s$ (pseudo-spin) given in terms of $e$-$e$ pair operators. The second symmetry was called "superconducting" by Yang because it is intimately connected with states that show long range order [2,5]. At each lattice site these symmetries are realized via **local generators**:

$$X^+_m = a^\dagger_\uparrow a_\downarrow, \quad X^-_m = (X^+_m)^\dagger, \quad H_m = n_\uparrow - n_\downarrow \quad \text{(magnetic rep.)}, \qquad (2)$$

$$X^+_s = a^\dagger_\uparrow a^\dagger_\downarrow, \quad X^-_s = (X^+_s)^\dagger, \quad H_s = n_\uparrow + n_\downarrow - 1 \quad \text{(superconducting rep.)}. \qquad (3)$$

The two sets of generators are related by the unitary transformation $a_\downarrow \leftrightarrow a^\dagger_\downarrow$. They generate mutually orthogonal algebras that are isomorphic to the algebra generated by the Pauli matrices and have unit elements (projectors)

$$1_s = H^2_s, \quad 1_m = H^2_m, \quad 1_s + 1_m = 1. \qquad (4)$$

The superconducting generators commute with each term of the local part of the Hubbard Hamiltonian
$$H^{(\text{loc})} = u \sum_i n_{i\uparrow} n_{i\downarrow} - \mu \sum_{i,\sigma} n_{i\sigma}, \tag{5}$$
provided that the chemical potential satisfies
$$\mu = \frac{u}{2} \quad \text{"half filling"}. \tag{6}$$
This can either be seen by direct computation or by studying the action of the generators on the four possible electron states ($\bigcirc, \text{\textcircled{$\downarrow$}}, \text{\textcircled{$\uparrow$}}, \text{\textcircled{$\updownarrow$}}$) at each site:

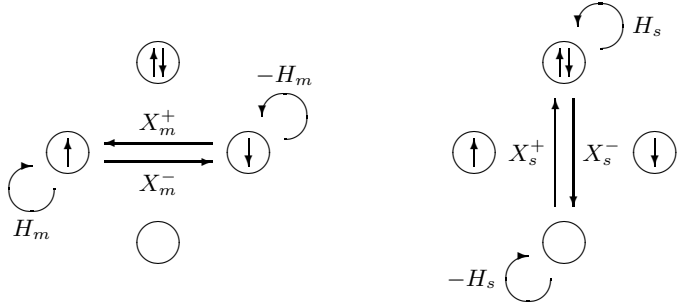

Consider *e.g.* the action of $[H^{(\text{loc})}, X_s^+]$ on the state "$\bigcirc$"; the result is "$\text{\textcircled{$\updownarrow$}}$" multiplied by $(u - 2\mu)$ and should vanish for symmetry (6). By similar computation it is also easily seen that the magnetic generators commute with each term of $H^{(\text{loc})}$—from now on we will however focus solely on the superconducting symmetry.

To check the symmetry of the non-local hopping term
$$H^{(\text{hop})} = t \sum_{\langle i,j \rangle \sigma} a_{j\sigma}^\dagger a_{i\sigma}, \tag{7}$$
we have to consider **global generators**: These generators act on the whole lattice and are given by the sum of the local generators for all sites. Here we are in effect taking tensor-products of representations; the abstract rule is given by the coproduct of $U(su(2))$.

The search for quantum group symmetries in the Hubbard model is motivated by the observation that the generators $X_s^+$, $X_s^-$ and $H_s$ in the superconducting representation of SU(2) also satisfy the SU$_q$(2) algebra as given in [6]
$$\left[X^+, X^-\right] = \frac{q^H - q^{-H}}{q - q^{-1}}, \qquad \left[H, X^\pm\right] = \pm 2 X^\pm. \tag{8}$$

(The proof uses $H_s^3 = H_s \Rightarrow q^{H_s} - q^{-H_s} = (q - q^{-1}) H_s$.) By quantum symmetry we mean invariance of the Hamiltonian under quantum adjoint action. This implies that the Hamiltonian commutes with all (global) generators and vice versa. It immediately follows that $H^{(\text{loc})}$ has a local quantum symmetry. Before we take the coproduct of the quantum group into account this is however still a trivial statement.

By a modification of the Hubbard Hamiltonian we now seek to extend the local symmetry to a non-trivial global quantum symmetry, defined via the coproduct
$$\Delta_q(X^\pm) = X^\pm \otimes q^{-H/2} + q^{H/2} \otimes x^\pm, \quad \Delta_q(H) = H \otimes 1 + 1 \otimes H \tag{9}$$

of $SU_q(2)$. The idea of [7] was to achieve this by including phonons. A Yangian version of such symmetries was previously considered in [8].

The **extended Hubbard model** of [7] (with some corrections as explained in [9]) introduces Einstein oscillators (parameters: $M$, $\omega$) and electron-phonon couplings (local: $\vec{\lambda}$-term, non-local: via $T_{ij}$):

$$H_{\text{ext}} = u \sum_i n_{i\uparrow} n_{i\downarrow} - \mu \sum_{i,\sigma} n_{i\sigma} - \vec{\lambda} \cdot \sum_{i\sigma} n_{i\sigma} \vec{x}_i$$
$$+ \sum_i \left( \frac{\vec{p}_i^{\,2}}{2M} + \frac{1}{2} M \omega^2 \vec{x}_i^{\,2} \right) + \sum_{\langle i<j \rangle \sigma} T_{ij} a^\dagger_{j\sigma} a_{i\sigma} + \text{H.C.} \qquad (10)$$

with hopping amplitude

$$T_{ij} = t \exp\left(-\hbar \zeta \hat{e}_{ij} \cdot \vec{\kappa}\right) \exp\left(\zeta \hat{e}_{ij} \cdot (\vec{x}_i - \vec{x}_j) + i\kappa \cdot (\vec{p}_i - \vec{p}_j)\right). \qquad (11)$$

The deflections $\vec{x}_i$ of the ions from their rest positions at each site and the corresponding momenta $\vec{p}_i$ satisfy canonical commutation relations. The $\hat{e}_{ij}$ are unit vectors from site $i$ to site $j$.

For $\vec{\kappa} = 0$ the model reduces to the Hubbard model with phonons and atomic orbitals in $s$-wave approximation [9].

**Local symmetry:** The local part $H^{(\text{loc})}_{\text{ext}}$ of the extended Hubbard Hamiltonian commutes with the generators of $SU_q(2)_s$

$$\left[ \tilde{X}^\pm_s, H^{(\text{loc})}_{\text{ext}} \right] = \left[ H_s, H^{(\text{loc})}_{\text{ext}} \right] = 0 \quad \text{iff} \quad \mu = \frac{u}{2} - \frac{\vec{\lambda}^2}{M\omega^2}. \qquad (12)$$

(The modified generators $\tilde{X}^\pm_s \equiv \exp(\mp \frac{2i\vec{\lambda}\cdot\vec{p}}{\hbar M\omega^2}) X^\pm_s$ also satisfy the $SU_q(2)$ algebra.)

**Global symmetry:** Through direct computation it is seen that the nonlocal part $H^{(\text{hop})}_{\text{ext}}$ of $H_{\text{ext}}$ and thereby the whole extended Hubbard Hamiltonian commutes with the global generators

$$\left[ \tilde{X}^\pm_s, H_{\text{ext}} \right] = [H_s, H_{\text{ext}}] = 0, \quad \text{iff} \quad \vec{\lambda} = \hbar M \omega^2 \vec{\kappa}, \quad q = \exp(2\kappa\zeta\hbar), \qquad (13)$$

where for $i < j$ next neighbor sites $\kappa \equiv -\hat{e}_{ij} \cdot \vec{\kappa}$. *For $q \neq 1$ the symmetry is restricted to models given on a 1-dimensional lattice.*

The extended Hubbard model shows a superconducting symmetry at $\mu \neq u/2$. With the help of a convenient change of basis we will now argue that we are however still at half filling: A **Lang-Firsov transformation** [10]

$$H_{\text{ext}} \to H_{q\text{-sym}} = U H_{\text{ext}} U^{-1}, \quad U = \exp(i\vec{\kappa} \cdot \sum \vec{p}_j n_{j\sigma}) \qquad (14)$$

leads to an equivalent Hamiltonian

$$H_{q\text{-sym}} = H^{(\text{loc})}_{\text{ext}}(\vec{\lambda}', u', \mu') + H^{(\text{hop})}_{q\text{-sym}}, \qquad (15)$$

that we shall call the *quantum symmetric Hubbard model*, with a new set of parameters $\vec{\lambda}'$, $u'$, $\mu'$ and a modified hopping term

$$H^{(\text{hop})}_{q\text{-sym}} = \tilde{t} \sum_{\langle i,j \rangle \sigma} a^\dagger_{j\sigma} a_{i\sigma} \underbrace{\left(1 + (q^{(i-j)/2} - 1) n_{i-\sigma}\right) \left(1 + (q^{(j-i)/2} - 1) n_{j-\sigma}\right)}_{\text{new}}. \qquad (16)$$

The coefficient $\tilde{t}$ is a function of $\vec{p}_i, \vec{p}_j$ but may also be turned into a temperature-dependent constant via a mean field approximation. This approximation is valid for the quantum symmetric Hubbard model because the condition for symmetry in the *new* parameters is easily seen to be

$$\vec{\lambda}' = 0, \qquad \mu' = \frac{u'}{2}, \tag{17}$$

*i.e.* requires vanishing local phonon coupling and corresponds to "half filling".

The mean field approximation does not break the quantum symmetric Hubbard model's superconducting symmetry, so phonons are not essential. For simplicity we will only consider $H_{q\text{-sym}}$ in mean field approximation, *i.e.* without phonons. In the remainder of the talk I will briefly sketch how the quantum symmetric model can be further simplified and in fact even related to the standard Hubbard model. The origin of the quantum symmetry turns out to be the *classical* SO(4) symmetry found in the standard Hubbard model.

**Extended Lang-Firsov transformation** We recall that the Hubbard model with phonons (with classical symmetry) can be related to the standard Hubbard Hamiltonian via two steps: A Lang-Firsov transformation that changes the model to one with vanishing local phonon coupling followed by a mean field approximation that removes the phonon operators from the model by averaging over Einstein oscillator eigenstates [11]. The question here is whether there exists a similar transformation that relates the quantum symmetric Hubbard model to the standard Hubbard model. It is easy to see that the hopping terms of $H_{q\text{-sym}}$ and $H_{\text{Hub}}$ have different spectrum so the transformation that we are looking for cannot be an equivalence transformation. There exists however an operator $M$, expressible in terms of (3) and (4), with $MM^* = 1 + k\xi$, $\xi^2 = \xi$ (*i.e.* similar to a partial isometry), that transforms the coproducts of the Chevalley generators $\bar{X}^\pm$, $\bar{H}$ into their quantum counterparts

$$M\Delta_c(\bar{X}^\pm)_s M^* = \Delta_q(X^\pm)_s, \quad M\Delta_c(\bar{H})_s M^* = \Delta_q(H)_s, \tag{18}$$

(the generators on the RHS are of Jimbo-Drinfel'd type) and the standard Hubbard Hamiltonian into $H_{q\text{-sym}}$

$$MH_{\text{Hub}}M^* = H_{q\text{-sym}}\ ! \tag{19}$$

Note however that $H_{\text{Hub}}$ and $H_{q\text{-sym}}$ are not equivalent because $M$ is not unitary. Still with the knowledge of the existence of $M$ the proof of the quantum symmetry of $H_{\text{ext}}$ is greatly simplified.

**Twists between quasi-Hopf algebras** A systematic way to study the relation of quantum and classical symmetries was given by Drinfel'd in his theory of quasi-Hopf algebras [12]. All we need to know here is that the undeformed U(su$_2$) and deformed U$_q$(su$_2$) are isomorph as algebras[1]

$$\text{U}(\text{su}_2) \overset{i}{\approx} \text{U}_q(\text{su}_2) \qquad i : \text{algebra isomorphism}, \tag{20}$$

and the classical/cocommutative ($\Delta_c$) and quantum ($\Delta_q$) coproducts are related via conjugation ("twist")

$$\Delta_q\left(i(x)\right) = \mathcal{F}\, i^{\otimes 2}(\Delta_c(x))\, \mathcal{F}^{-1}, \qquad \mathcal{F} \in \text{U}_q(\text{SU}(2))^{\hat{\otimes} 2}. \tag{21}$$

---
[1] More precisely: $\text{U}(\text{su}_2)[[h]] \overset{i}{\approx} \text{U}_q(\text{su}_2)$, where $h = \ln q$.

Let us focus on a pair of next neighbor sites on the lattice. We are interested in a representation of the universal $\mathcal{F}$ on the Hilbert space of states of two sites:

$$F = (\epsilon_m \oplus \rho_s)^{\otimes 2}(\mathcal{F}) = 1 \otimes 1 - 1_s \otimes 1_s + F_s. \qquad (22)$$

The operator $F_s \equiv \rho_s^{\otimes 2}(\mathcal{F})$ is obtained from the known spin-$\frac{1}{2}$ representation of the $F$-matrix by plugging in (3).

We now face a puzzle: By construction $F^{-1}H_{q\text{-sym}}F$ should commute with the (global) generators of $\mathrm{SU}_q(2)_s$ just like $H_{\mathrm{Hub}}$—but it cannot be equal to $H_{\mathrm{Hub}}$ because of the different spectrum. The resolution of this puzzle is that there exists actually whole families both of classically and quantum symmetric models. The quantum symmetric hopping term (16) is for instance a special case of

$$\sum_{\langle i<j \rangle \sigma} a_{j-\sigma}^\dagger a_{i-\sigma} \left(1 + (q^{\frac{1}{2}}c - 1)n_{i\sigma}(1 - n_{j\sigma}) + (q^{-\frac{1}{2}}c^* - 1)(1 - n_{i\sigma})n_{j\sigma}\right) + \text{H.C.}, \qquad (23)$$

where $c$ is a free complex parameter. All these models are related by extended Lang-Firsov transformations and classically and quantum symmetric models come in twist-equivalent pairs.

Explicit expressions for $M$, $F$ and all Hubbard Hamiltonians involved have been found but their derivation and presentation are beyond the scope of this talk. To close I would nevertheless like to give a schematic picture of the complete family of classically *i.e.* $\mathrm{SU}(2)_s \times \mathrm{SU}(2)_m/\mathbf{Z}$-symmetric extended Hubbard Hamiltonians. The most general Hamiltonian with this symmetry has six free real parameters:

$$H_{\text{sym}} = \text{BOSONIC TERMS} + \text{HOPPING TERM} + \text{QUANTUM HOPPING TERM} \qquad (24)$$

The bosonic terms include $\mu \sum_{\sigma i} n_{\sigma i} - u \sum_i n_{\uparrow i} n_{\downarrow i}$, a spin-wave term, a pair-hopping term with coefficient $2\mu - u$ (and hence related to the filling factor!) and density-density interaction terms. The hopping term is $\sum_{\langle i,j \rangle \sigma} a_{\sigma i}^\dagger a_{\sigma j} + h.c.$; the quantum hopping term leads by twisting to the particular hopping term deformation that we found in $H_{q\text{-sym}}$. Depending on the coefficient of the pair-hopping term some of the models based on $H_{\text{sym}}$ show symmetry away from half filling. The known and many (all) other quantum symmetric Hubbard models are obtained from $H_{\text{sym}}$ by (multi-parametric) twists as described above. While twisting provides additional parameters, the advantage of the classical models obtained from $H_{\text{sym}}$ for specific choices of parameters is that their symmetry is not restricted to one dimension—that is very important for more realistic models of the conduction planes in cuprates:

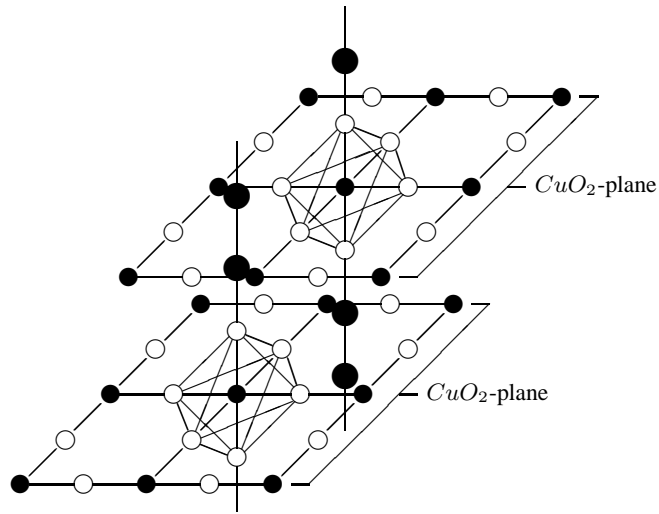

## Acknowledgments

It is a pleasure to thank Michael Schlieker, Nicolai Reshetikhin and Bruno Zumino for crucial input into this work, Julius Wess for ongoing support, B. L. Cerchiai for early collaboration and Gaetano Fiore for suggestions concerning the manuscript.